\documentstyle[12pt,epsfig,twoside]{article}

\newcommand{\be}{\begin{equation}}
\newcommand{\ee}{\end{equation}}
\newcommand{\dlt}{\delta}
\newcommand{\Dlt}{\Delta}

\newcommand{\vp}{\varphi}
\newcommand{\bt}{\beta}
\newcommand{\al}{\alpha}
\newcommand{\prt}{\partial}

\newcommand{\om}{\omega}

\newcommand{\ep}{\varepsilon}

\newcommand{\br}{{\bf r}}

\begin{document}

\begin{center}

{\Large{\bf Dynamics of Nonground-State Bose-Einstein Condensates} \\ [5mm]

V.I. Yukalov$^1$ and E.P. Yukalova$^2$} \\ [3mm]

{\it $^1$Bogolubov Laboratory of Theoretical Physics, \\
Joint Institute for Nuclear Research, Dubna 141980, Russia \\ [2mm]
$^2$Department of Computational Physics, Laboratory of Information 
Technologies, \\
Joint Institute for Nuclear Research, Dubna 141980, Russia}

\end{center}

\vskip 1cm

\begin{abstract}

Dilute Bose gases, cooled down to low temperatures below the
Bose-Einstein condensation temperature, form coherent ensembles
described by the Gross-Pitaevskii equation. Stationary solutions
to the latter are topological coherent modes. The ground state,
corresponding to the lowest energy level, defines the standard
Bose-Einstein condensate, while the states with higher energy
levels represent nonground-state condensates. The higher modes
can be generated by alternating fields, whose frequencies are
in resonance with the associated transition frequencies. The
condensate with topological coherent modes exhibits a variety
of nontrivial effects. Here it is demonstrated that the dynamical
transition between the mode-locked and mode-unlocked regimes is
accompanied by noticeable changes in the evolutional entanglement
production.

\end{abstract}

\vskip 5mm


\vskip 2cm

\section{Introduction}

\begin{sloppypar}
Bose-Einstein condensation, as is known, occurs when a macroscopic
number of bosons piles down to the same lowest single-particle state.
In equilibrium, this is the sole possible type of Bose-Einstein
condensate. When the Bose-condensed system is rotated, vortices can
appear. The Bose-Einstein condensate with vortices is the most known
example of a nonground-state condensate. At low
temperatures Bose gases are described by the Gross-Pitaevskii equation
(see reviews [1--3]). In the presence of a trapping potential, there
arises a whole spectrum of energy levels associated with the stationary
solutions to the Gross-Pitaevskii equation. Each of these stationary
solutions is, by definition, a topological coherent mode [4]. The lowest
energy level corresponds to the standard Bose-Einstein condensate, while
the higher states describe various nonground-state condensates. The latter
can be generated by alternating fields, whose frequencies should be in
resonance with the related transition frequencies [4--8]. It is feasible
to realize  an oscillatory modulation of either the trapping potential or
of the atomic scattering length. The trapping potential can be either
single-well or multiwell.

Nonground-state Bose-Einstein condensates, with generated topological
coherent modes, exhibit several interesting effects, such as interference
fringes, interference current, mode locking, dynamical transition, critical
phenomena, chaotic motion, harmonic generation, parametric conversion,
and atomic squeezing [4--8]. The aim of the present communication is to
study the behaviour of entanglement production in a trapped nonground-state
condensate with topological coherent modes and to demonstrate that the
properties of this entanglement production experience noticeable changes
under the dynamical transition from the mode-locked regime to mode-unlocked
regime.

\end{sloppypar}

\section{Topological Coherent Modes}

Dilute Bose-condensed gases at low temperature are described [1--3] by
a coherent field $\vp(\br,t)$ satisfying the Gross-Pitaevskii equation
\be
\label{1}
i\hbar\; \frac{\prt}{\prt t} \; \vp(\br,t) =
\left (\hat H[\vp]+\hat V\right )\; \vp(\br,t) \; ,
\ee
in which $\hat V=\hat V(\br,t)$ is a modulating potential and the nonlinear
Hamiltonian
\be
\label{2}
\hat H[\vp] = -\hbar^2 \; \frac{\nabla^2}{2m} + U(\br) +
N\; A_s \; |\vp(\br,t)|^2
\ee
contains a trapping potential $U(\br)$ and the interaction intensity
$A_s\equiv 4\pi \hbar^2 a_s/m$ with $m$ being atomic mass; $a_s$,
scattering length; and $N$, the total number of atoms.

The topological coherent modes are the solutions to the stationary
Gross-Pitaevskii equation
\be
\label{3}
\hat H[\vp_n]\; \vp_n(\br) = E_n\; \vp_n(\br) \; ,
\ee
where $n$ is a labelling multi-index. The generation of topological modes
is accomplished by transferring atoms from an occupied mode (say, ground
state) to a higher mode by means of an alternating field
\be
\label{4}
V(\br,t) = \frac{1}{2}\; B(\br) e^{i\om t} + \frac{1}{2}\; B^*(\br)
e^{-i\om t} \; ,
\ee
whose frequency $\om$ is in resonance with a chosen transition frequency
$\om_{21}\equiv(E_2-E_1)/\hbar$ between two selected energy levels, implying
that the detuning $\Dlt\om\equiv\om-\om_{21}$ is small, $|\Dlt\om|\ll\om$.
Generally, it is admissible to apply several resonant fields of form (4)
connecting different modes. The procedure of such a resonant mode generation
has been thoroughly explained [4--8].

The solution to Eq. (1) is expressible as
\be
\label{5}
\vp(\br,t) = \sum_n c_n(t)\; \vp_n(\br) \; \exp\left ( -\;
\frac{i}{\hbar}\; E_n\; t\right ) \; ,
\ee
where $c_n(t)$ are functions of time to be found. Substituting expansion
(5) into Eq. (1) makes it possible to obtain the equations for $c_n(t)$.
The dynamics of two selected modes, connected by the resonant field (4),
is described by the temporal dependence of the corresponding coefficients
$c_1(t)$ and $c_2(t)$, whose moduli squared, $|c_n|^2$, define the
fractional mode populations. Let us introduce the population difference
$s\equiv|c_2|^2-|c_1|^2$ and the phase difference $x\equiv i\ln(c_1|c_2|/
|c_1|c_2)$.

The dynamical properties of a given atomic system, prepared in a state
with fixed initial conditions, can be governed by varying the amplitude
$|\bt|\equiv(\vp_1,\hat B\vp_2)/\hbar$ and the detuning $\Dlt\om
\equiv\om-\om_{21}$ of the modulating field (4), with $(\vp_1,\vp_2)$
being a scalar product. It is convenient to deal with dimensionless
parameters normalizing $|\bt|$ and $\Dlt\om$ by the value $\al\sim
A_sN/\hbar$ characterizing the atomic interaction strength. Then one
deals with the parametric manifold $\{ b,\dlt\}$, in which $b\equiv
|\bt|/\al$ is the dimensionless modulating-field amplitude and $\dlt$
is the dimensionless detuning. On this parametric manifold, there is
a critical line $\{ b_c,\dlt_c\}$ separating two qualitatively different
regimes of motion. From one side of the critical line, there occurs the
{\it mode-locked regime}, when the fractional mode populations are locked
in their initial half-planes, such that for all times $t$ one has either
$0\leq|c_n(t)|^2\leq 1/2$ or $1/2\leq|c_n(t)|^2\leq 1$, depending on
the initial value $c_n(0)$. And from another side of the critical line,
the mode populations become unlocked, oscillating in the whole region
$0\leq|c_n(t)^2\leq 1$, which corresponds to the {\it mode-unlocked
regime}. In this way, by varying either $b$ or $\dlt$, and crossing
the critical line $\{ b_c,\dlt_c\}$, one can realize the {\it dynamic
transition} from one regime of motion to another, between the mode-locked
and mode-unlocked regimes. On the plane of the variables $s$ and $x$,
the dynamic transition happens when the initial point of a trajectory
crosses a saddle separatrix given by the equation $s^2/2-b\sqrt{1-s^2}
\cos x +\dlt s=|b|$. Thus in Fig. 1 the initial point $\{-1,0\}$, for
the parameter $b<b_c$, is below the saddle separatrix, because of which
the total related trajectory lies below the separatrix, which corresponds
to the mode-locked regime of motion. In Fig. 2 the same initial point,
but for $b>b_c$, is already above the separatrix, which results in the
mode-unlocked regime.

\vskip 5mm
{\bf Figure 1}: {\it Phase portrait on the plane$\{ s,x\}$ for a two-mode
condensate, with the parameters $\dlt=-0.1$, $b=0.35$; $b_c=0.39821$}.

\vskip5mm

{\bf Figure 2}: {\it Phase portrait on the plane$\{ s,x\}$ for a two-mode
condensate, with the parameters $\dlt=-0.1$, $b=0.49$; $b_c=0.39821$}.

\section{Evolutional Entanglement Production}

The dynamic transition, with changing regimes of motion, should also be
accompanied by noticeable changes in all dynamical characteristics of the
system, for instance, in the behaviour of atomic squeezing [6]. Here we
concentrate on the temporal behaviour of entanglement production in the
considered system. Entanglement is an important characteristic that is
assumed to be exploited for quantum information processing and quantum
computation. We consider here the entanglement generated by the
second-order density matrix $\hat\rho_2$ for the given coherent two-mode
condensate. The entanglement generated by $\hat\rho_2$ can be described
[9,10] by the entanglement-production measure $\ep_2(t)\equiv
\ep(\hat\rho_2)=\log(||\hat\rho_2||_D/||\hat\rho_2^\otimes||_D)$,
in which the logarithm is to base $2$, $\hat\rho_2^\otimes$ is the
nonentangling counterpart of $\hat\rho_2$ and $D$ is the set of
disentangled states formed of the products $\vp_m(\br_1)\vp_n(\br_2)$.
For the density matrix $\hat\rho_2$, we obtain
$\ep_2(t)=-\log\sup_n|c_n(t)|^2$.

Numerical calculations for the entanglement-productions measure $\ep_2(t)$
are presented in Figs. 3 and 4 for the initial conditions $c_1(0)=1$ and
$c_2(0)=0$. Fixing the zero detuning, we vary the modulating-field amplitude
$b$. In Fig. 3, the latter is below the critical value $b_c$, while in
Fig. 4, it is above $b_c$. As is seen, the entanglement production is rather
different in the mode-locked regime, when $b<b_c$, and in the mode-unlocked
regime, when $b>b_c$. The considered entanglement, generated in a trap,
quantifies the amount of correlations between each pair of atoms. If atoms
are released from the trap, say in the time-of-flight regime, then the
studied measure $\ep_2(t)$ will change in time, with the entanglement
produced in a trap serving as the initial condition. Depending on the
latter, the pair correlations of outcoupled atoms will be different and
could be regulated by preparing the required entanglement production of
trapped atoms. A two-mode system is mathematically similar to two-level
atoms or to $1/2$ spins. Therefore such a system can also be employed
for realizing quantum gates, which can be done by switching on and off
the external field. Thus, by varying the amplitude of the modulating field
one can essentially modify the characteristics of entanglement production,
which could be employed for information processing.

\begin{figure}[ht]
\centerline{\psfig{file=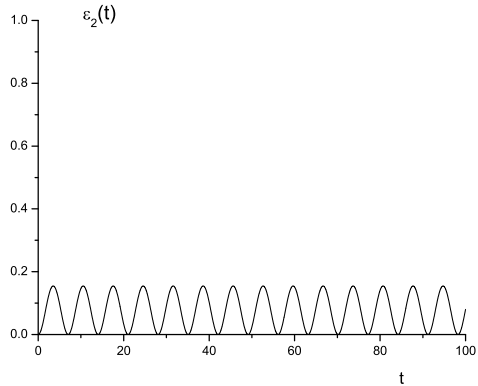,height=2in}}
\caption{Evolution of the entanglement produced by the second-order
density matrix for $\dlt=0$, $b=0.3$. Time is measured in units of $\al$.}
\label{fig:Fig.3} \end{figure}

\begin{figure}[ht]
\centerline{\psfig{file=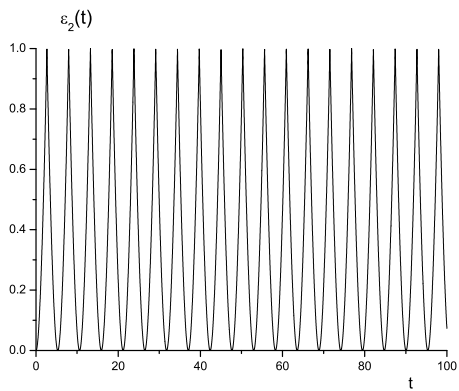,height=2in}}
\caption{Entanglement production of the second-order density matrix as 
a function of time, measured in units of $\al$, for $\dlt=0$, $b=0.7$.}
\label{fig:Fig.4} \end{figure}

In conclusion, the nonground-state Bose-Einstein condensates, with
resonantly generated topological coherent modes, are the objects whose
dynamical properties can be straightforwardly regulated by external
modulating fields. By means of such fields, it is feasible to switch
the dynamic behaviour between two qualitatively different regimes, the
mode-locked and mode-unlocked  regimes. This change-over from one regime
to another allows one to radically alter the system evolution, with the
related changes in the behaviour of the fractional mode populations, atomic
squeezing, and entanglement production.





\end{document}